\documentclass[twocolumn,amssymb,superscriptaddress,aps,prd,floatfix]{revtex4}
\usepackage{amsmath}
\usepackage[dvips]{graphicx}
\usepackage[dvips]{color}
\usepackage[usenames,dvipsnames]{xcolor}
\newlength{\colw}
\renewcommand{\Re}{\operatorname{Re}}

\newcommand{\One}{1\kern-4.5pt1}

\definecolor{DarkGreen}{rgb}{0,0.5,0}

\begin{document}

\title{Lattice Gluon and Ghost Propagators, and the Strong Coupling in Pure SU(3) Yang-Mills Theory: Finite Lattice Spacing and Volume Effects}

\author{Anthony G. Duarte}
\author{Orlando Oliveira}
\author{Paulo J.\ Silva}
\affiliation{CFisUC, Department of Physics, University of Coimbra, P--3004 516 Coimbra, Portugal.}

\begin{abstract}
The dependence of the Landau gauge two point gluon and ghost correlation functions on the lattice spacing and on the physical volume
are investigated for pure SU(3) Yang-Mills theory in four dimensions using lattice simulations.
We present data from very large lattices up to $128^4$ and for two lattice spacings $0.10$ fm and $0.06$ fm corresponding to volumes
of $\sim$ (13 fm)$^4$ and $\sim$ (8 fm)$^4$, respectively.  
Our results show that, for sufficiently large physical volumes, both propagators
have a mild dependence on the lattice volume. 
On the other hand, the gluon and ghost propagators change with the
lattice spacing $a$ in the infrared region, with the gluon propagator having a stronger dependence on $a$ compared to the
ghost propagator. 
In what concerns the strong coupling constant $\alpha_s (p^2)$, as defined from gluon and ghost two point functions,  
the simulations show a sizeable dependence on the lattice spacing for the infrared region and for momenta up to $\sim 1$ GeV.
\end{abstract}

\pacs{11.15.Ha,12.38.Aw,21.65.Qr}
         
\maketitle

\section{Introduction}

The computation of the gluon and ghost propagators of pure Yang-Mills theory in the Landau gauge
have been investigated in the past years using lattice simulations for the SU(2) and SU(3) groups
and accessing deeper the infrared region. 
In four dimensions, this effort established a consensus that the gluon
propagator~\cite{Cucchieri:2007md,Cucchieri:2007zm,Bogolubsky:2009dc,Dudal:2010tf,Cucchieri:2011ig,Ilgenfritz:2010gu,Oliveira:2010xc,Oliveira:2012eh,Sternbeck:2012mf}
is infrared suppressed and acquires a finite non-vanishing value at zero momentum.
On the other hand, the ghost propagator seems to be described essentially by its tree level 
expression~\cite{Sternbeck:2005vs,Oliveira:2006zg,Cucchieri:2007zm,Cucchieri:2008fc,Bogolubsky:2009dc,Ilgenfritz:2010gu,Cucchieri:2013nja,Cucchieri:2016jwg}. In order to access the infrared momenta, the lattice simulations have been performed on huge volumes: (27 fm)$^4$ using a $128^4$ lattice
for the SU(2) gauge group~\cite{Cucchieri:2007md} and (17 fm)$^4$ using a $96^4$ lattice for SU(3). Such large volumes were achieved
by setting the lattice spacing at $\sim 0.2$ fm. Although the simulations were performed within the perturbative scaling window, the use of
such large lattice spacings rise the question of how far from the continuum limit the results are. 

For pure Yang-Mills theory the mass scale is given by the mass of the lightest glueball state.
For SU(3), lattice simulations~\cite{Chen:2005mg,Chowdhury2015}  show that the lightest glueball has the quantum numbers
$J^{PC} = 0^{++}$ and a mass of about $1700$ MeV. The distance scale associated with such a value for the mass 
being $a \sim 0.12$ fm. For the SU(2) gauge group~\cite{Teper98}, 
the predicted lightest glueball is about $1600$ MeV and the corresponding distance scale is given again by $0.12$ fm. 
It follows, that the large lattice spacings used in the simulation mentioned in the previous paragraph can introduce some bias on the final
result. Indeed, in~\cite{Oliveira:2012eh}, the dependence of the gluon propagator on the lattice spacing for the
SU(3) gauge group was analysed. The authors conclude that not only the effects due to the use of a large lattice spacing are dominant,
over the finite volume effects, but also that the computations using such a  large lattice spacing underestimates the propagator in the 
infrared region. However, at the qualitative level the results of~\cite{Oliveira:2012eh} reproduce the large volume/large lattice spacing
simulations reported in~\cite{Cucchieri:2007md,Bogolubsky:2009dc}.

In the current paper, we aim to extend the work of~\cite{Oliveira:2012eh} and investigate the dependence of the gluon and 
ghost propagators on the lattice spacing for large physical volumes $\gtrsim 6.5$ fm. 
Furthermore, given that from the gluon and ghost propagator one can define a renormalisation group invariant
strong coupling constant $\alpha_s (p^2)$, we also analyse the dependence of the coupling on the lattice spacing. Our results
show that the use of a large lattice spacing changes the deep infrared values of the gluon propagator, of the ghost propagator
and of the strong coupling constant. The simulations reported here show that the gluon propagator is suppressed in the infrared region,
when one uses a large lattice spacing, while the ghost propagator is enhanced by using a larger lattice spacing.
On the other hand, for the definition of the strong coupling constant considered here, the use of a larger lattice spacing enhances 
$\alpha_s$ for low and mid momenta up to $p \lesssim 1$ GeV.

The paper is organised as follows. In Sec.~\ref{Sec:Setup} we resume the details of the lattice calculations, including definitions,
number of configurations,  Landau gauge fixing and the renormalization procedure. In Sec.~\ref{SubSec:Gluon} we report
on the computation of the gluon propagator, while in Sec.~\ref{Sec.Ghost} we report on the results for the ghost propagator.
In Sec.~\ref{Sec:coupling} the results for the running coupling are discussed. In Sec.~\ref{Sec:compbma} we compare our simulatons
with the lattice results of~ \cite{Bogolubsky:2009dc}. Finally, in Sec.~\ref{Sec:fim} we summarise the results discussed and conclude.

\section{Lattice Setup and Renormalization Procedure \label{Sec:Setup}}

The pure gauge SU(3) Yang-Mills simulations reported here use the Wilson action at several $\beta$ values and physical volumes. The full set of 
simulations performed in the context of this work is resumed in Tab.~\ref{tab:setup}. 
For the conversion into physical units we use the string tension as measured in~\cite{Bali1993}. 

The gauge configurations were generated with the Chroma library~\cite{Edwards2005} using a combined Monte Carlo sweep of seven overrelaxation
updates with four heat bath updates. 
Each configuration $U_\mu (x)$ obtained from the Monte Carlo sampling was gauge fixed to the Landau gauge by maximising the functional
\begin{equation}
   F_U[g] = \frac{1}{V \, N_d \, N_c} \sum_{x , \mu} \Re\mbox{Tr} \left[ g(x) \, U_\mu (x) \, g^\dagger( x + \hat{e}_\mu) \right]
   \label{GFfunctional}
\end{equation}   
over the gauge orbit and where $V$ is the number of the lattice points, $N_d = 4$ the number of space-time dimensions, $N_c = 3$ the number
of colours and $\hat{e}_\mu$ the unit vector along the direction $\mu$. 
In what concerns the gauge fixing algorithm, we rely on the Fourier accelerated steepest descent method~\cite{Davies1988}, which was implemented using
Chroma and PFFT~\cite{Pippig2013} libraries. The quality of the gauge fixing was monitored by
\begin{equation}
\theta = \frac{1}{V \, N_c} \sum_x \mbox{Tr}\left[ \Delta (x) \, \Delta^\dagger (x) \right] \ ,
\end{equation} 
where
\begin{equation}
 \Delta (x)  = \sum_\nu \left[  \, U_\nu (x - \hat{e}_\mu ) - U_\nu (x) - h.c. - \mbox{ trace } \right] \ ,
\end{equation} 
which is the lattice version of the gauge fixing condition $\partial A = 0$. For each gauge configuration, the gauge fixing was stopped when
$\theta \leq 10^{-15}$.

\begin{table}[t]
   \centering
   \begin{tabular}{l@{\hspace{0.25cm}}c@{\hspace{0.25cm}}l@{\hspace{0.25cm}}r@{\hspace{0.25cm}}r@{\hspace{0.25cm}}cc} 
      \hline 
      \hline
      $\beta$   & $a$ (fm)  & $1/a$ (GeV)   & $L$ & $La$ (fm) &Conf & Sources \\
      \hline
       5.7      & 0.1838(11) & 1.0734(63)  & 44   &  8.087    &   100   &    3   \\
       6.0      & 0.1016(25) & 1.943(47)   & 64   &  6.502    &   100   &    2   \\
                &            &             & 80   &  8.128    &    70   &    2   \\
                &            &             & 128  & 13.005    &    35   &    1   \\
       6.3      & 0.0627(24) &  3.149(46)  & 128  &  8.026    &    54   &    3   \\
       \hline
       \hline
   \end{tabular}
   \caption{Lattice setup. The physical scale was set from the string tension as measured by~\cite{Bali1993}. 
                The lattice spacing for $\beta = 6.3$ was not measured in~\cite{Bali1993}, so we relied on
                the procedure described in~\cite{Silva2014}. The last column refers to the number of point sources, per configuration, used 
                in the inversion of the Faddeev-Popov matrix needed to compute the ghost propagator. }
   \label{tab:setup}
\end{table}

The Landau gauge gluon propagator is given by
\begin{equation}
   D^{ab}_{\mu\nu} (p) = \delta^{ab} \left( \delta_{\mu\nu} - \frac{p_\mu p_\nu}{p^2}\right) \, D(p^2) \ ,
\end{equation}   
where latin indexes refer to colour degrees of freedom and greek indexes to Lorentz degrees of freedom,
and its computation was done using the set of definitions described in Ref.~\cite{Silva2004}. 
The results reported here are given as function of the tree level improved momentum
\begin{equation}
 p_\mu = \frac{2}{a} \sin \left( \frac{n \, \pi}{L_\mu} \right) \, , \qquad  n = 0, 1, \dots, \frac{L_\mu}{2}
\end{equation}
where $a$ is the lattice spacing and $L_\mu$ the number of lattice points in the direction $\mu$. 
The statistical errors on the propagators were evaluated using the bootstrap method with a confidence level of 67.5\%.

The ghost propagator is defined as
\begin{equation}
   G^{ab} (p) = \delta^{ab} \, G(p^2)
\end{equation}
and we have relied on the method described in~\cite{Suman1996} to compute the scalar function $G(p^2)$.
For most of the ensembles, we have considered several sources and
their results averaged, in order to reduce the statistical noise.
The statistical errors for the ghost propagator were computed as for the gluon propagator.

In the current paper, besides the propagators we also look at the renormalization group invariant strong coupling defined by
\begin{equation}
   \alpha_s (p^2) = \frac{g^2_0}{4 \pi} \, d_D(p^2) \, d^2_G (p^2) \, ,
\end{equation}
where 
\begin{equation}
  d_D(p^2) = p^2 \, D(p^2) \quad\mbox{ and }\quad d_G (p^2)  = p^2 \, G(p^2)
  \label{Eq:dressing}
\end{equation}
are the gluon and ghost dressing functions, respectively.

In order to compare the data of the various simulations, the propagators were renormalized using a MOM scheme
with the renormalized propagators defined as
\begin{equation}
 \left. D(p^2) \right|_{p^2 = \mu^2} = Z_A \, D_{Lat}(\mu^2)= \frac{1}{\mu^2}
 \label{Eq:GlueZ}
\end{equation} 
and
\begin{equation}
 \left. G(p^2) \right|_{p^2 = \mu^2} = Z_\eta \, G_{Lat}(\mu^2)= \frac{1}{\mu^2} 
 \label{Eq:GhostZ} 
\end{equation} 
where $D_{Lat}$ and $G_{Lat}$ refer to the bare lattice propagators. In the current work we use $\mu = 4$ GeV
for the renormalization scale. The renormalization constants $Z_A$ and $Z_\eta$ were computed by fitting the bare lattice 
propagators to the functional form
\begin{equation}
    D(p^2)  =   z ~\frac{p^2 + m^2_1}{p^4 + m^2_2 \, p^2 + m^4_3} \label{Eq:FitRenGlue}
\end{equation}
in the range $p \in [ 0 \, , \, 6]$ for the gluon propagator and
\begin{equation}
    G(p^2)  =  z  \, \frac{ \left[  \log \frac{p^2}{\Lambda^2} \right]^{\gamma_{gh}} } {p^2}  \label{Eq:FitRenGhost}
\end{equation}
in the range $p \in [ 2 \, , \, 6]$ for the ghost propagator. Then, we use the fits to impose the normalization conditions (\ref{Eq:GlueZ}) and (\ref{Eq:GhostZ}).
We have checked that the fits reproduced the lattice data for momentum $\sim 4$ GeV.  Furthermore, in all cases the $\chi^2/d.o.f.$ associated to the fits
are below unit.

In order to reduce the lattice artefacts, for momenta above 1 GeV we have performed the conic cut as defined in~\cite{Leinweber1998}.
For momenta below 1 GeV, the figures include all lattice data points. 

Besides the finite size effects due to the simulation on a finite box,
with a finite lattice spacing, any lattice calculation is imbued with Gribov noise. In the current work we do not attempt to estimate the
effects coming from the choice of the various maxima of the functional (\ref{GFfunctional}). As discussed in e.g.~\cite{Silva2004, Silva2007, Sternbeck2013},
peaking different maxima of $F_U[g]$ can lead to small changes in the propagators in the infrared region.

\section{Propagators and Strong Coupling: how they change with the lattice spacing and the volume \label{Sec:Gluon}}

\begin{figure*}[t]
   \centering
   \includegraphics[scale=0.36]{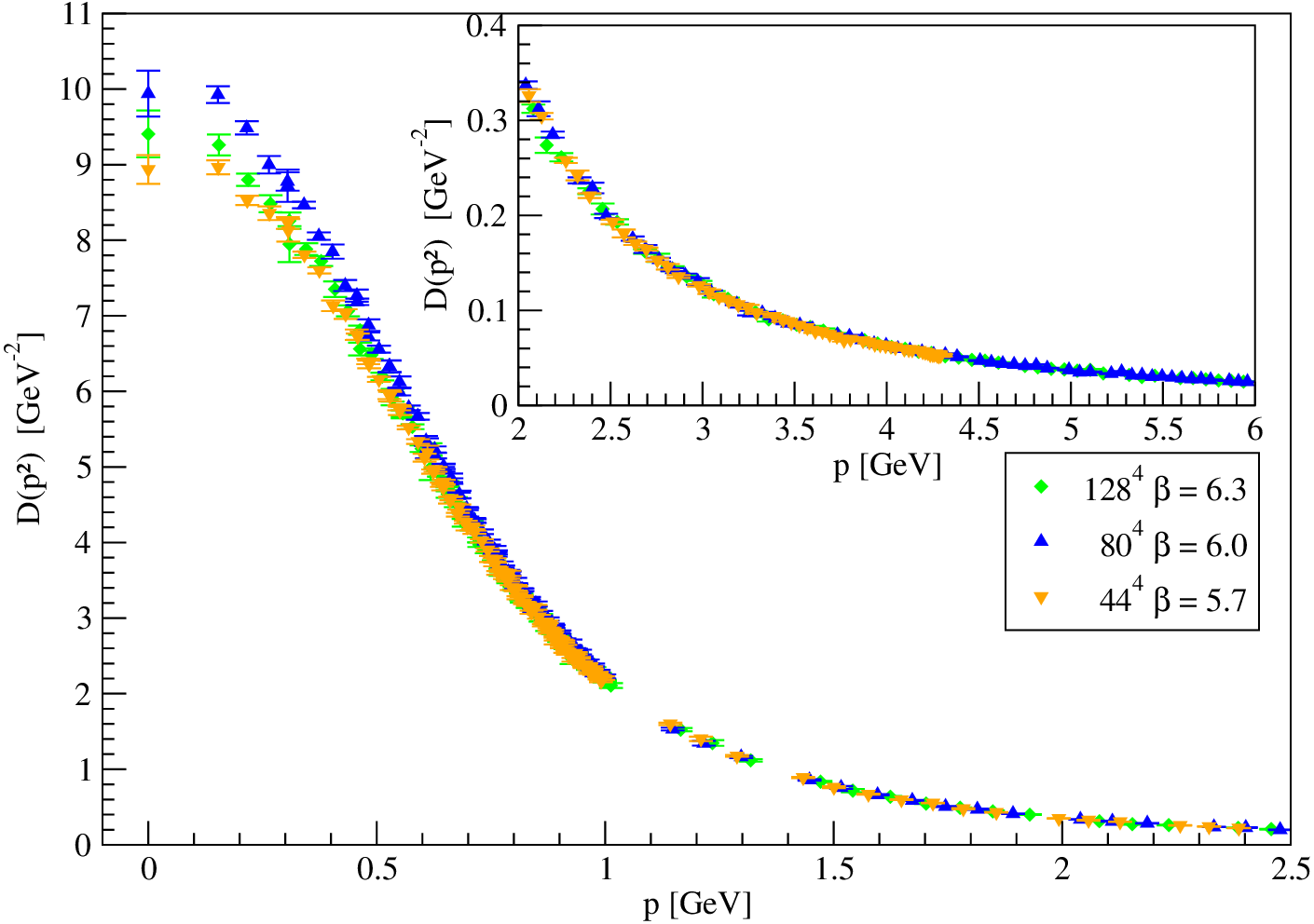} 
   \hspace*{2mm}
   \includegraphics[scale=0.36]{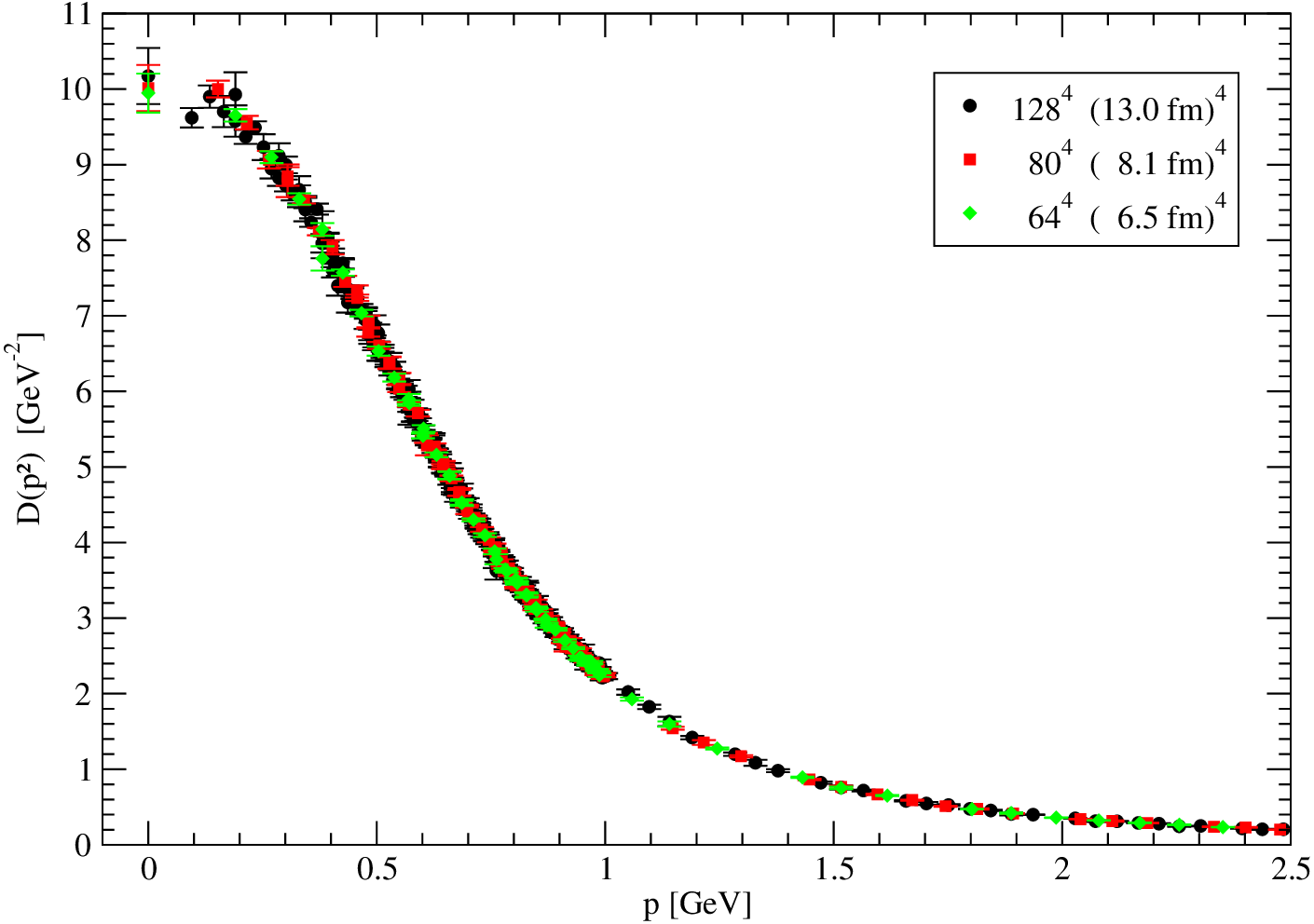} 
   \caption{Renormalised gluon propagator at $\mu = 4$ GeV for: (\textit{left}) a physical volume of (8 fm)$^4$ 
                 and different lattice spacings;   (\textit{right}) the same lattice spacing and different volumes. 
                 Details about the lattice parameters are given in Tab.~\ref{tab:setup}. }
   
   \label{fig:gluon_prop_4GeV}
\end{figure*}

In this section we present the results of the simulations resumed in Tab.~\ref{tab:setup}, focusing on the dependence on
the lattice spacing and physical lattice volume.

\subsection{The Gluon Propagator \label{SubSec:Gluon}}

The data for the renormalised gluon propagator can be seen in figure~\ref{fig:gluon_prop_4GeV}. 
In the left plot, the lattice data for essentially the same volume ($V \sim 8$ fm) and
different lattice spacings (0.18 fm, 0.10 fm and 0.063 fm) is compared, whereas the right plot  
outlooks the simulations performed with the same lattice spacing ($ a \sim 0.10$ fm) 
but different physical volumes ($La =$ 6.5 fm, 8.1 fm and 13.0). If the data of our simulations
shows essentially no dependence on the physical volume for volumes above (6.5 fm)$^4$, 
it also reveals a non-trivial dependence of the propagator
on the lattice spacing. 

From Fig.~\ref{fig:gluon_prop_4GeV}  one concludes that for the same physical volume, using a larger lattice spacing has an
impact on the gluon propagator for momenta up to $\lesssim 1$ GeV, with the larger lattice spacing underestimating the 
lattice data in the infrared region.

The relative importance of finite lattice spacing/finite volume effects confirms the results reported in~\cite{Oliveira:2012eh}.

\begin{figure*}[t]
   \vspace*{7mm}
   \centering
   \includegraphics[scale=0.36]{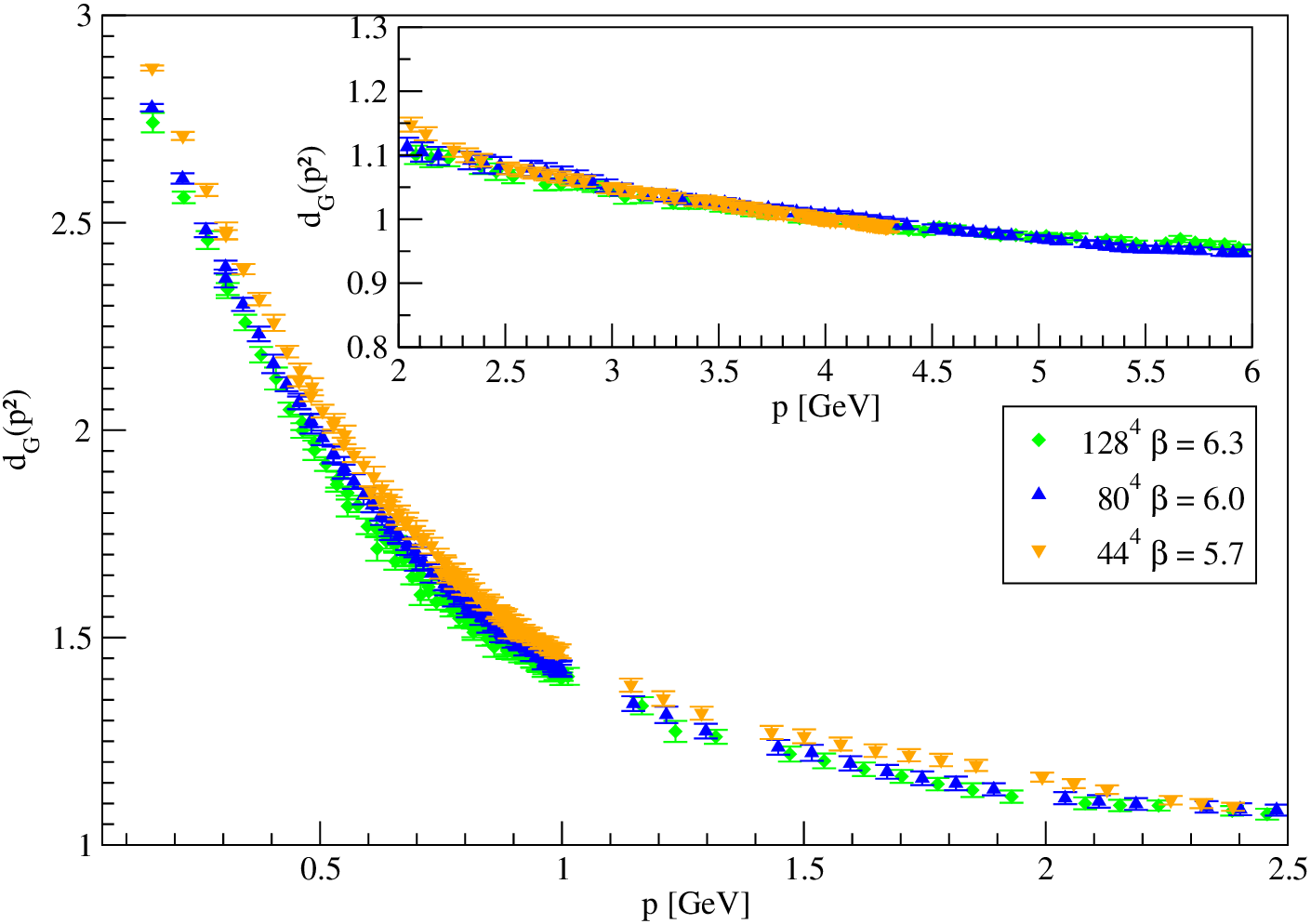} 
   \hspace*{2mm}
   \includegraphics[scale=0.36]{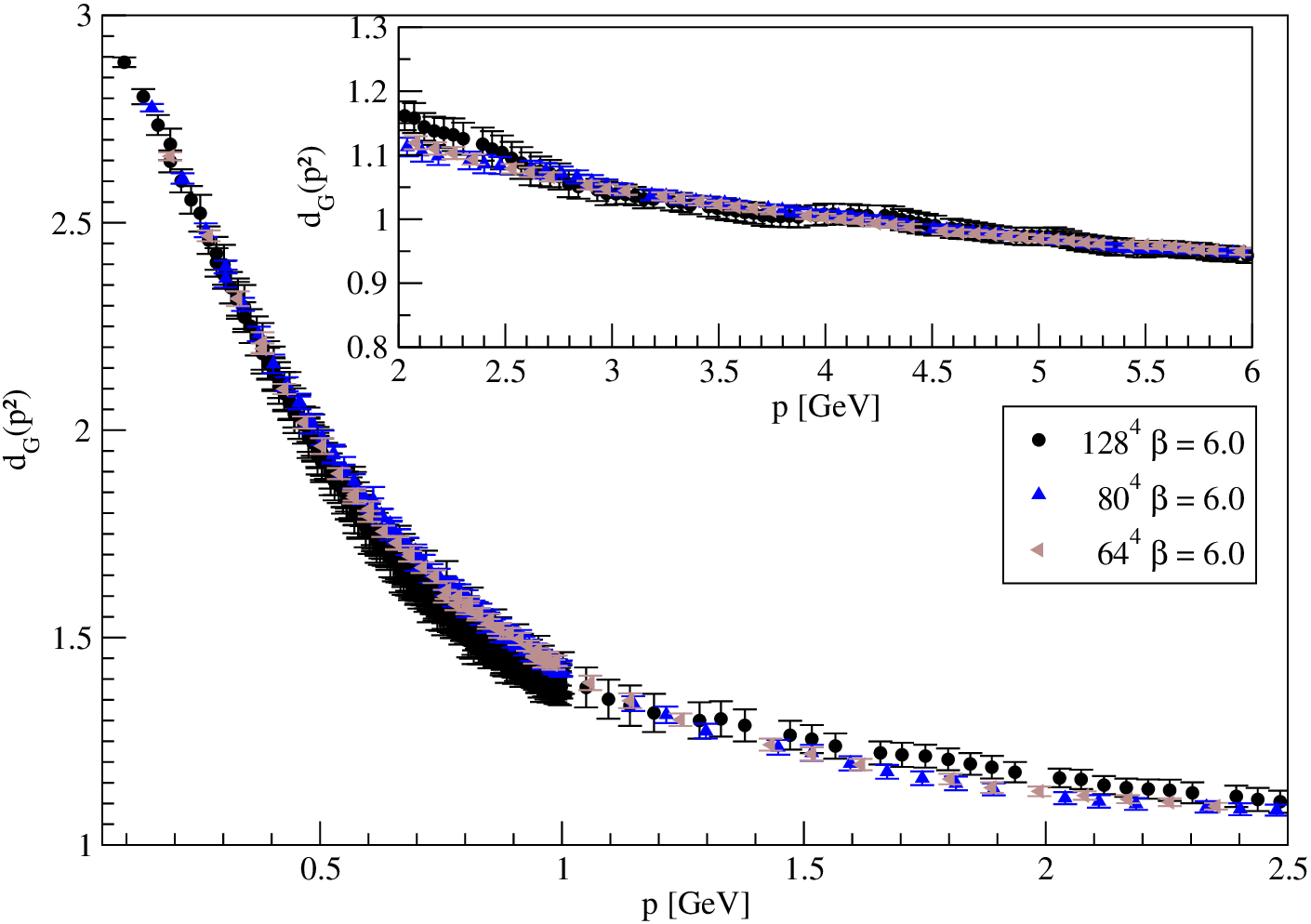} 
   \caption{Renormalised ghost dressing function at $\mu = 4$ GeV for:  (\textit{left}) a physical volume of (8 fm)$^4$ and different lattice spacings;
                  (\textit{right}) for the same lattice spacing and different volumes. Details about the lattice parameters 
                 are given in Tab.~\ref{tab:setup}.}
   \label{fig:ghost_dressing_4GeV}
\end{figure*}

\begin{figure}[t]
   \centering
   \includegraphics[scale=0.36]{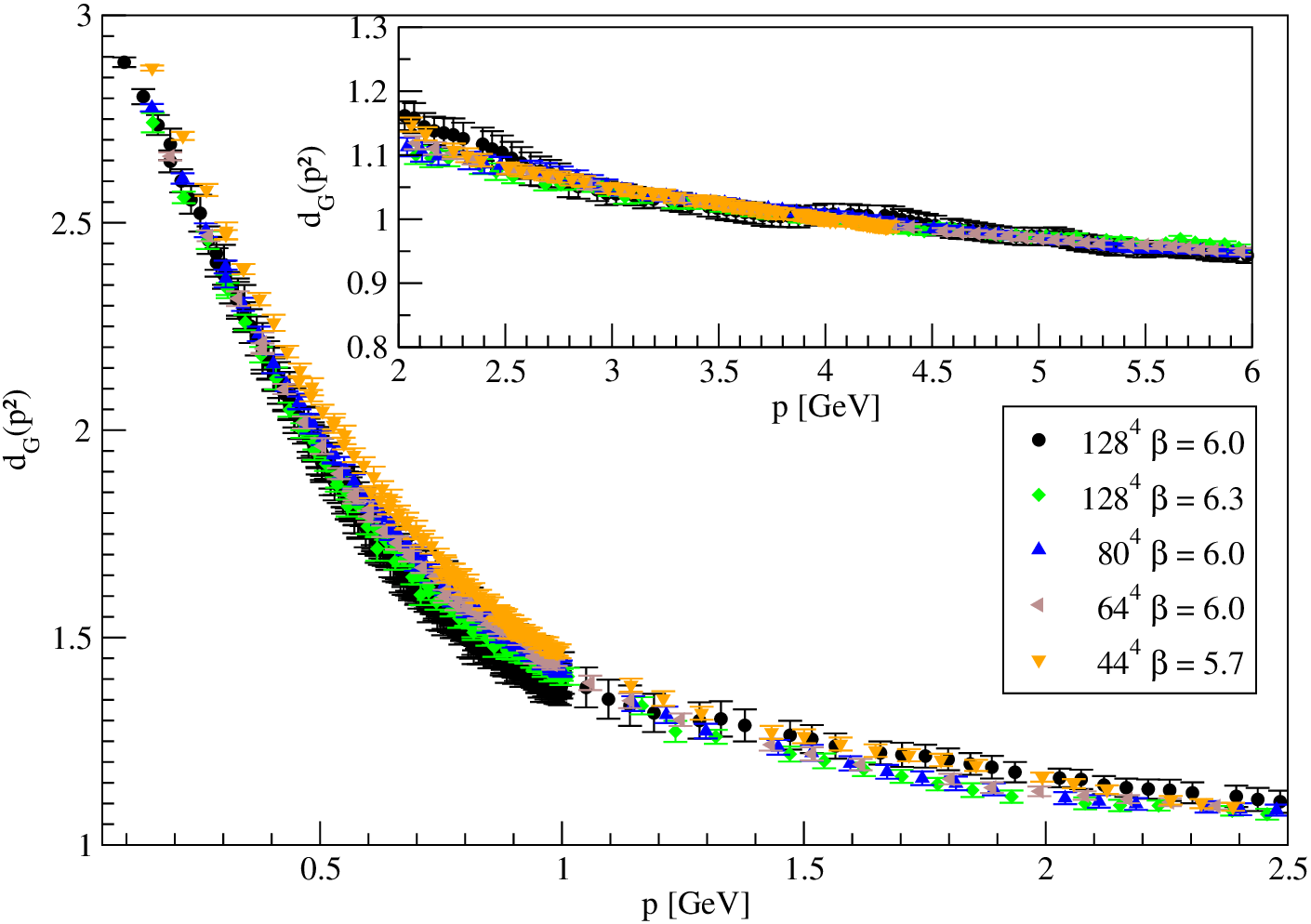} 
   \caption{Renormalised ghost dressing function at $\mu = 4$ GeV for the simulations reported in Tab.~\ref{tab:setup}.}
   \label{fig:ghost_dressing_prop_all}
\end{figure}

\subsection{The Ghost Dressing Function \label{Sec.Ghost}}

For the ghost two point function we report on the dressing function $d_G (p^2)$ as defined in Eq.~(\ref{Eq:dressing}). 
The ghost dressing function for the simulations 
with a physical volume of about (8 fm)$^4$ (left plot) and the same lattice spacings 
but different physical volumes (right plot) can be seen in  Fig.~\ref{fig:ghost_dressing_4GeV}. 

In what concerns the dependence on the lattice spacing, the figure shows that decreasing 
the lattice spacing, while keeping the same physical volume, suppresses the ghost propagator 
in the infrared region. The figure also shows that the data computed with our coarser lattice, 
i.e. the simulation performed with $\beta = 5.7$, differs from all the other simulations 
for momenta as large as 2 GeV. Indeed, for momenta up to 2 GeV, the $\beta = 5.7$ data 
is above the data of remaining simulations and, in this sense, the coarser lattice provides 
an upper bound to the corresponding continuum correlation function. Recall that the behaviour 
of the gluon propagator is the opposite, i.e. the $\beta = 5.7$ gives a lower bound to the 
continuum gluon propagator. The results of the simulations with the smaller lattice
spacings are compatible within one standard deviation only for momenta above $\sim 1$ GeV and
in the infrared region the propagator is suppressed if the lattice spacing is decreased. 
Note, however, that within two standard deviations the
dressing functions are compatible for almost the full range of momenta.

From the right panel of Fig.~\ref{fig:ghost_dressing_4GeV} one can conclude that, as for the gluon propagator, the dependence of the
lattice data on the physical volume is very mild if any. Indeed, the three lattice simulations are compatible within one standard deviation for all
momenta. The data for our largest physical volume has a larger statistical error, and seems not to be as smooth as the others, but this is possibly due to the
limited statistical ensemble used in the calculation of the correlation function.

For completeness, in Fig.~\ref{fig:ghost_dressing_prop_all} we report on the ghost dressing function for all the simulations refered
in Tab.~\ref{tab:setup}. The data for all the simulations agree within two standard deviations, with the exception of the $\beta = 5.7$ for a lattice
using $44^4$ points which overestimates the propagator.

\begin{figure*}[t]
   \centering
   \includegraphics[scale=0.36]{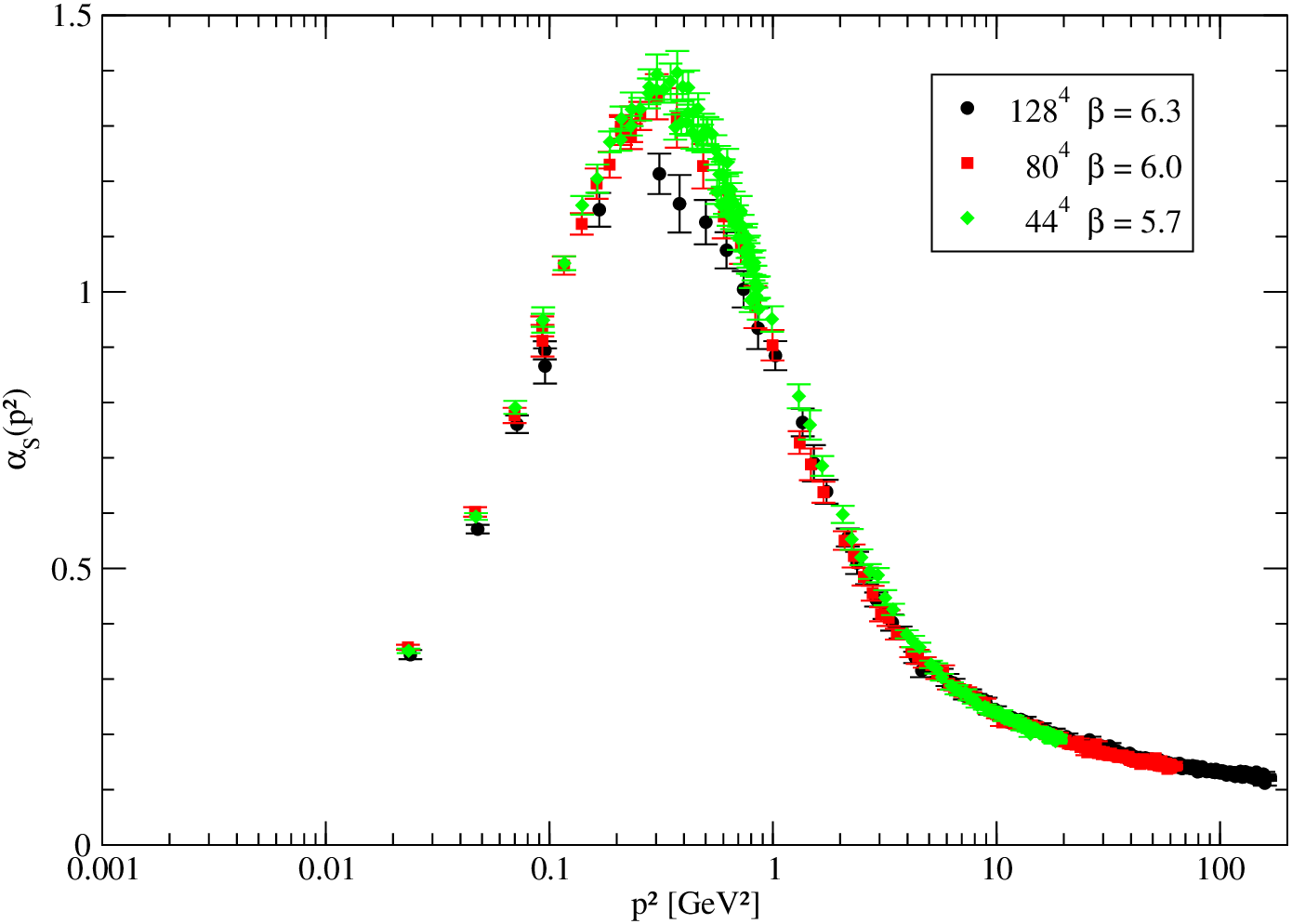} 
   \hspace*{3mm}
   \includegraphics[scale=0.36]{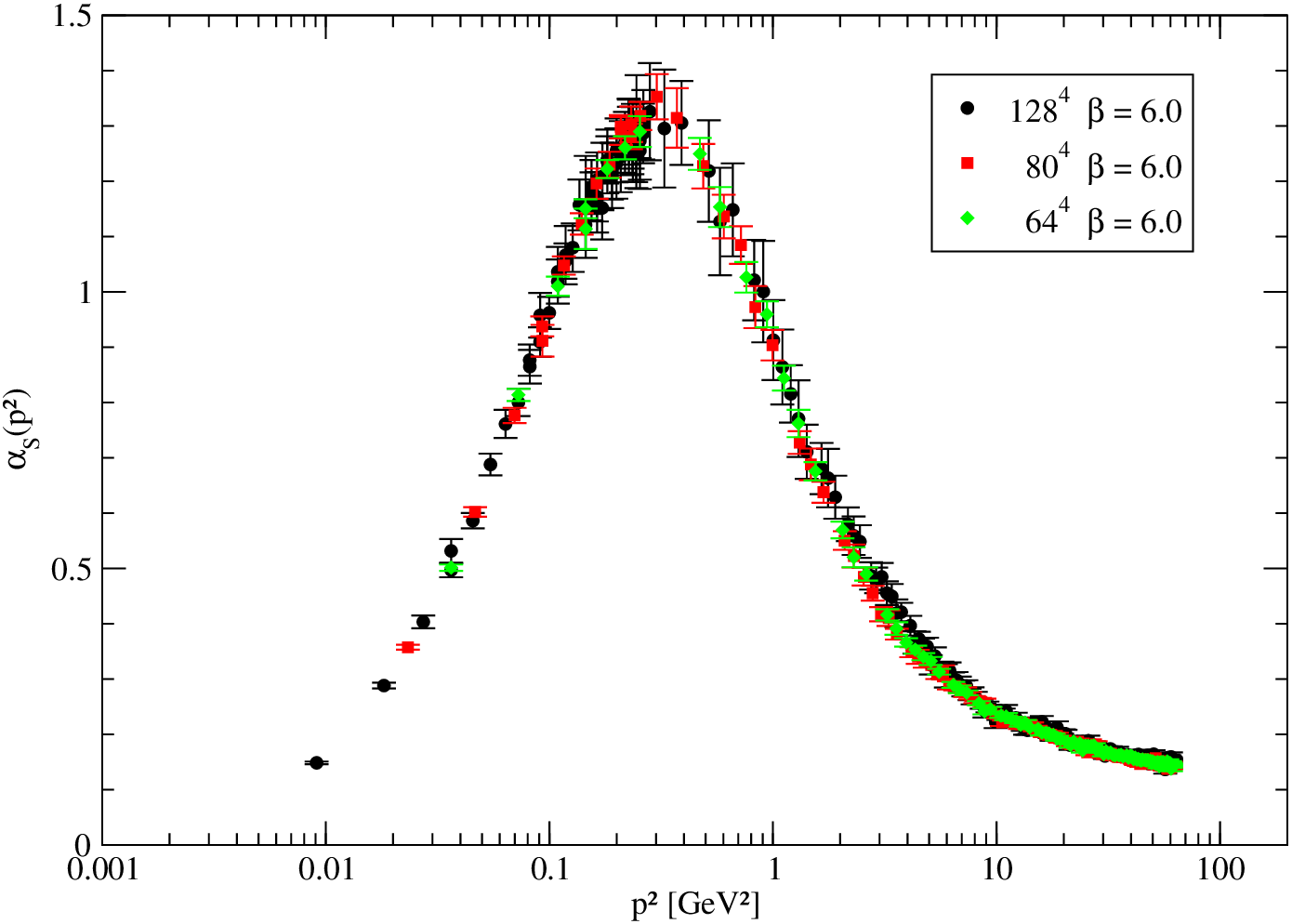} 
   \caption{Running coupling at $\mu = 4$ GeV for a physical volume of (8 fm)$^4$ 
and different lattice spacings (\textit{left}) and for the same lattice spacing and 
different volumes (\textit{right}). Details about the lattice parameters
 are given in Tab.~\ref{tab:setup}.}
   \label{fig:alpha_comp}
\end{figure*}

\subsection{The Running Coupling \label{Sec:coupling}}

The combination of dressing functions
\begin{equation}
\alpha_s (p^2) = \frac{g^2_0}{4 \pi} \, d_D (p^2) \, d^2_G(p^2)
\end{equation}
is a renormalization group invariant and defines a running coupling. 
As for the propagators, we also aim to understand how
$\alpha_s (p^2)$ changes with the lattice spacing and volume. 
In the computation of the strong coupling constant, we have used the bare lattice functions.

The dependence of the strong coupling on the lattice spacing and physical volume 
is resumed in Fig.~\ref{fig:alpha_comp}. In order to better illustrate the dependence on the physical volume and
lattice spacing, for the strong coupling constant, the plots only include the data surviving the momentum cuts mentioned before.
As can be observed from the right plot,
the simulations show essentially no dependence on the physical volume. 
On the other hand, the left plot of Fig.~\ref{fig:alpha_comp} shows that, at low and mid momenta, i.e. 
for $p \lesssim 1$ GeV, the strong coupling constant $\alpha_s (p^2)$ is slightly suppressed for smaller lattice spacings. 
For momenta above $\sim 1$ GeV, the results of all the simulations become compatible.

Another feature of $\alpha_s (p^2)$ concerns the position of its maximum. Indeed, as can be seen in
Fig.~\ref{fig:alpha_comp}, the position of the maximum of the strong coupling constant, as a function of $p^2$, seems to be independent
of both the lattice spacing and physical volume and occurs for $p^2 \sim 250$ MeV$^2$. However, in what concerns the numerical value of the maximum of $\alpha_s (p^2)$,
its value seems to be  suppressed when approaching the continuum limit, i.e. for smaller lattice spacings.
Indeed, our simulation at $\beta = 6.3$ shows a maximum of $\alpha_s(p^2)$ which is about 15\% smaller compared to the 
corresponding value obtained for the remaining simulations.

\section{Comparison with previous works \label{Sec:compbma}}

In this section we aim to compare our lattice results with those performed using the, so far, largest physical volumes for an
SU(3) simulation~\cite{Bogolubsky:2009dc}. We call the reader's attention that in this work, the conversion into physical units
relied on a different definition of the lattice spacing. In order to be able to compare these results with those reported in the
previous sections, we have rescaled the propagators accordingly. Another issue that should be taken into consideration is
that our simulations and those of~\cite{Bogolubsky:2009dc} used completely different algorithms to maximise the functional~(\ref{GFfunctional}).
As discussed previously, the choice of the maxima of $F_U[g]$ has an impact on the propagators, changing their behaviour in the infrared
region (Gribov noise) and, therefore, the comparison of the results at low momenta should be done with care. 

In Table~\ref{tab:berlinsetup} we summarise the lattice setup of the Berlin-Moscow-Adelaide simulations when one relies on our definition
for the conversion into physical units.

\begin{table}[t]
   \centering
   \begin{tabular}{l@{\hspace{0.25cm}}c@{\hspace{0.25cm}}l@{\hspace{0.25cm}}r@{\hspace{0.25cm}}r@{\hspace{0.25cm}}cc} 
      \hline 
      \hline
      $\beta$   & $a$ (fm)  & $1/a$ (GeV)   & $L$ & $La$ (fm) & \multicolumn{2}{c}{\# Conf}  \\
                     &                 &                        &       &                 & Glue   & Ghost \\
       \hline
       5.7         & 0.1838(11) & 1.0734(63)  & 64   & 11.763   & 14 & 14\\
                     &                   &                    &  72   & 13.234  & 20 & --\\
                     &                   &                    &  80   & 14.704  & 25 & 11\\
                     &                   &                    &  88   &  16.174 & 68 & --\\
                     &                   &                    &  96   &  17.645 & 67 & --\\
      \hline
      \hline
   \end{tabular}
   \caption{ Lattice setup considered by the Berlin-Moscow-Adelaide group ~\cite{Bogolubsky:2009dc}. Note that the data 
has been rescaled to use the same definition for all simulations --- see text for details.}
   \label{tab:berlinsetup}
\end{table}

\begin{figure}[t]
   \vspace*{5mm}
   \centering
   \includegraphics[scale=0.36]{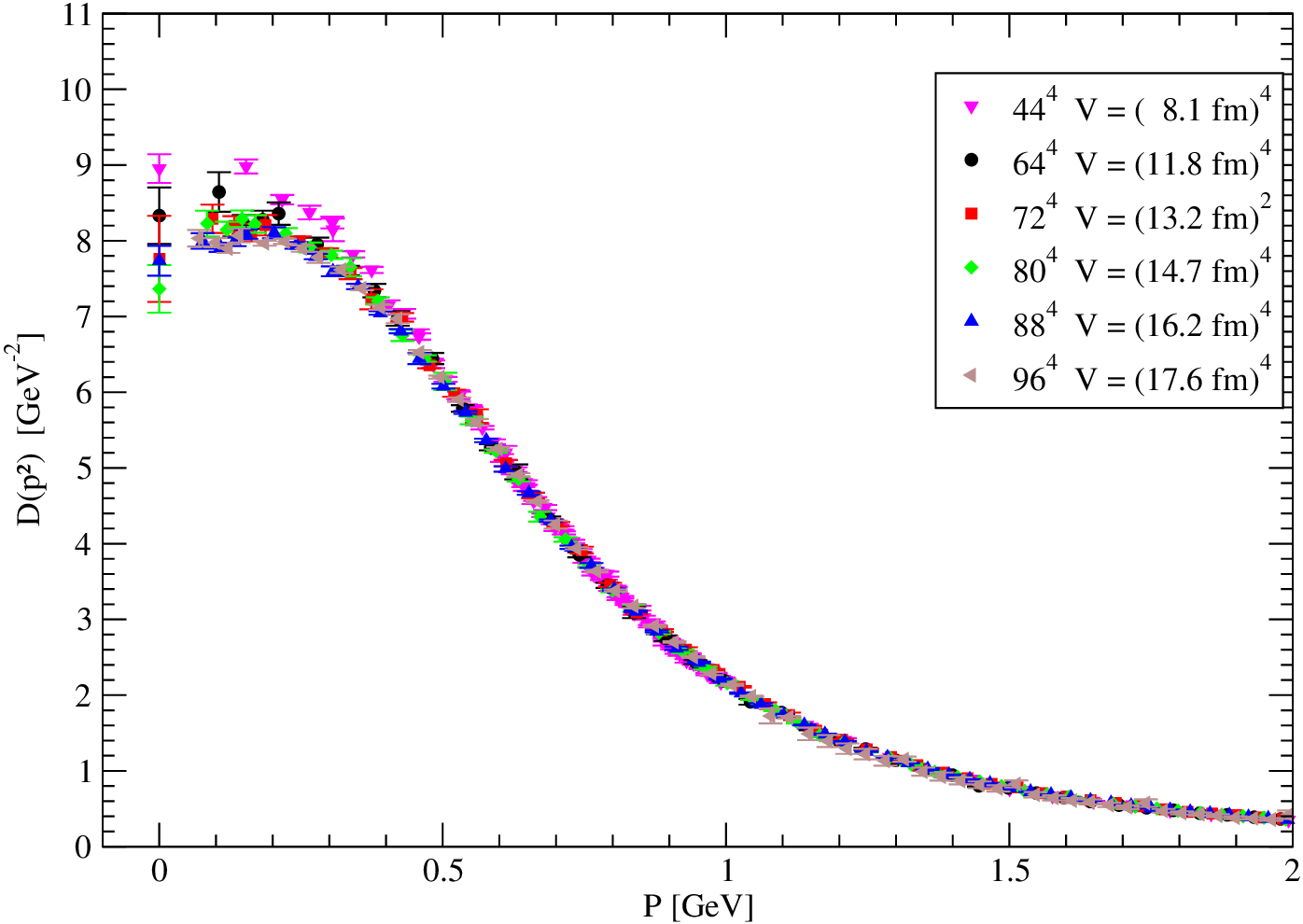} 
   \caption{Renormalized gluon propagator for the Berlin-Moscow-Adelaide lattice data. The plot also includes the results of our simulation with
                the same $\beta$ value.}
   \label{fig:glue_prop_berlin_renormalised}
\end{figure}

\begin{figure}[t]
   \vspace*{5mm}
   \centering
   \includegraphics[scale=0.36]{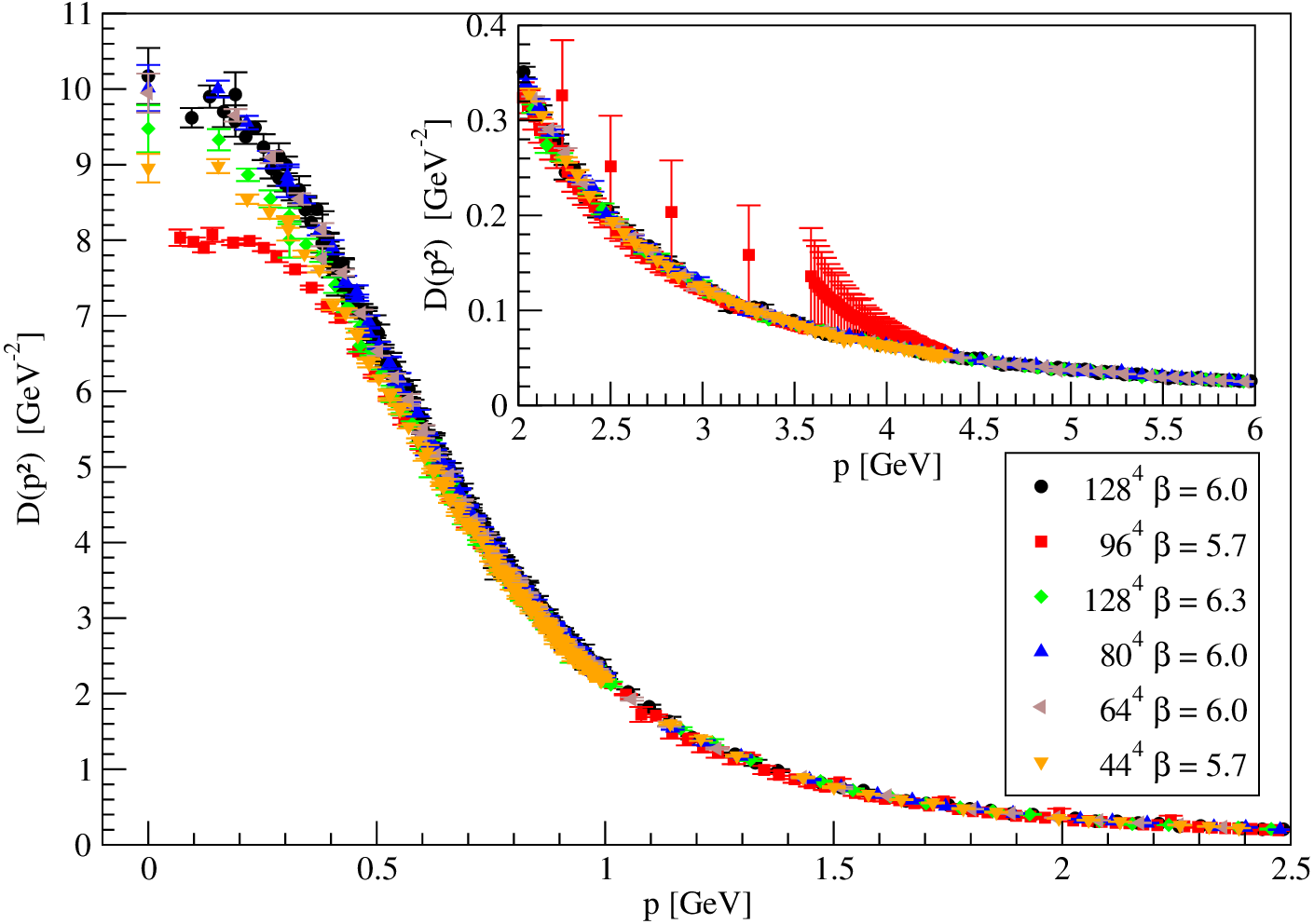} 
   \caption{All data sets including the largest volume of the Berlin-Moscow-Adelaide group. The inset shows a close-up of the high-momentum region,
                 where all data sets define a unique curve.}
   \label{fig:gluon_prop_all}
\end{figure}

\subsection{The Gluon Propagator}

In Fig.~\ref{fig:glue_prop_berlin_renormalised} we gather the results of our simulation at $\beta=5.7$
with those of the Berlin-Moscow-Adelaide group. 
The data shows that the dependence on the volume is at most mild, with the infrared propagator decreasing slightly 
when $La$ changes from 8 fm to 17 fm. Note that the differences occur only for momenta below $\sim 400$ MeV.

In Fig.~\ref{fig:gluon_prop_all} the two point gluon correlation function data reported previously, i.e. using larger (smaller) 
values of $\beta$ (the lattice spacing) but smaller physical volumes, is compared with the largest volume result of
the Berlin-Moscow-Adelaide group.  All data sets seems to converge into a unique curve for momenta above $\sim 0.7$ GeV.
For smaller momenta, the lattice data coming from the simulations at $\beta = 5.7$, which have the largest lattice spacing,
are always below the remaining results. The comparison of the simulations performed at the smallest $\beta$ values suggests
that the propagators associated with the higher $\beta$ should be multiplied by $\sim 8/9$, in the infrared region, to obtain the infinite volume limit.

\subsection{The Ghost Dressing Function}

\begin{figure}[t]
   \centering
   \includegraphics[scale=0.36]{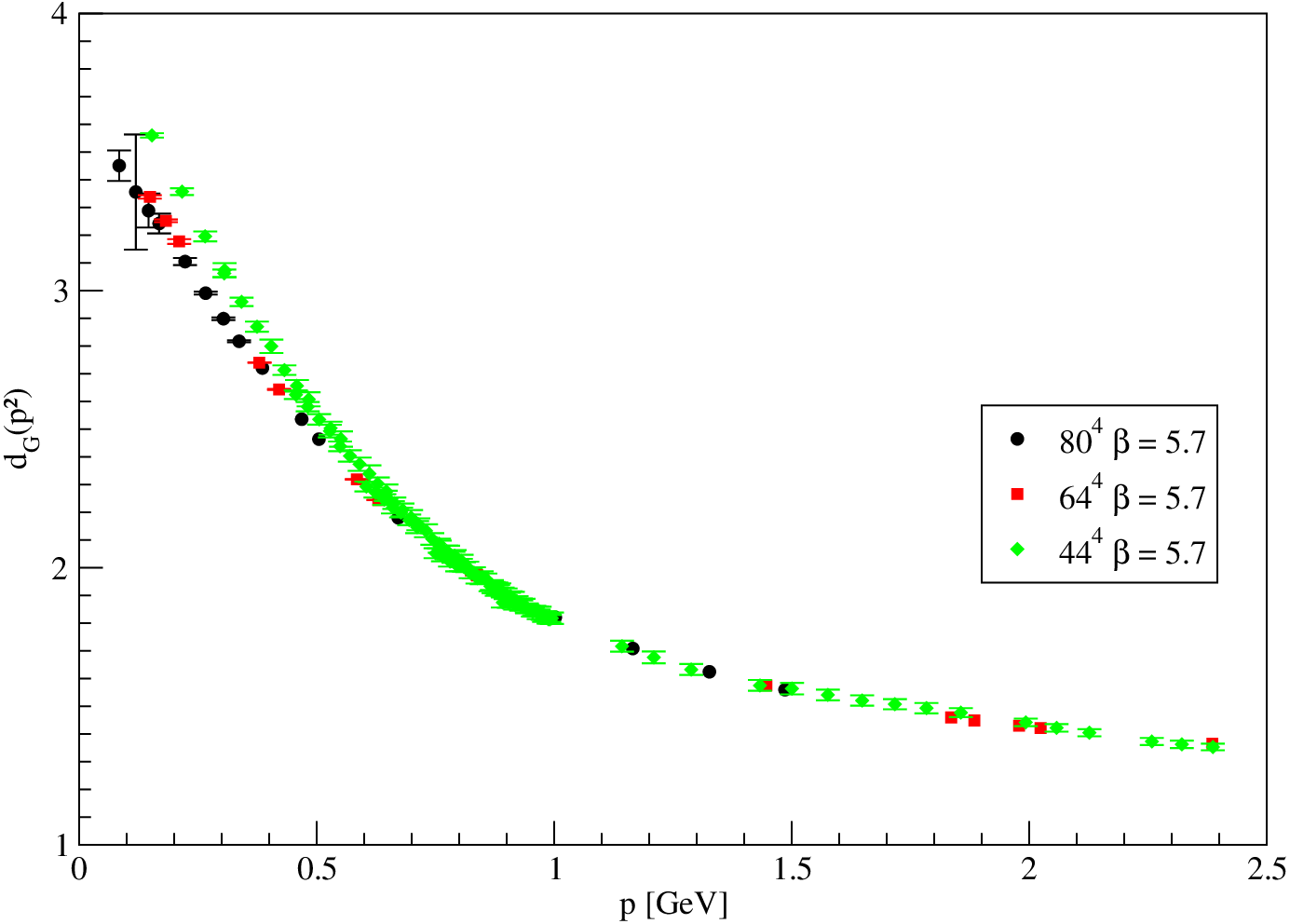} 
   \caption{Bare ghost dressing function from $\beta = 5.7$ simulations. The smallest lattice 
volume was rescaled to reproduce the $64^4$ Berlin-Moscow-Adelaide numbers at its largest momentum.}
   \label{fig:ghost_dressing_prop_Coimbra_Berlin_renormalised}
\end{figure}

The Berlin-Moscow-Adelaide ghost data covers momenta up to $\sim 3$ GeV ($\beta = 5.7$, $64^4$) or
up to $\sim 1.5$ GeV for the larger volume ($\beta = 5.7$, $80^4$). 
Given that we are considering as renormalization scale $\mu = 4$ GeV, one can not rescale the Berlin-Moscow-Adelaide
data to compare with our simulations, as was done for the gluon propagator.

In Fig.~\ref{fig:ghost_dressing_prop_Coimbra_Berlin_renormalised} our data for the ghost dressing function obtained
for $\beta = 5.7$ is compared with the results of the Berlin-Moscow-Adelaide collaboration. 
The bare lattice data from the $44^4$ lattice simulation was rescaled to reproduce the $64^4$ data at the highest available momentum.
It follows that for momenta above $\sim 700$ MeV, the results of the various simulations define a unique curve. On the other hand,
for smaller momenta the dressing function decreases as the physical volume of the lattice increases. 
This type of behaviour with the volume is not observed in our simulations where we used smaller lattice spacings. 
Indeed, as resumed in Fig.~\ref{fig:ghost_dressing_4GeV}, our data shows essentially no dependence on the physical volume in the infrared region.

\subsection{The Running Coupling \label{Sec:BMAcoupling}}

\begin{figure}[t]
   \vspace*{7mm}
   \centering
   \includegraphics[scale=0.36]{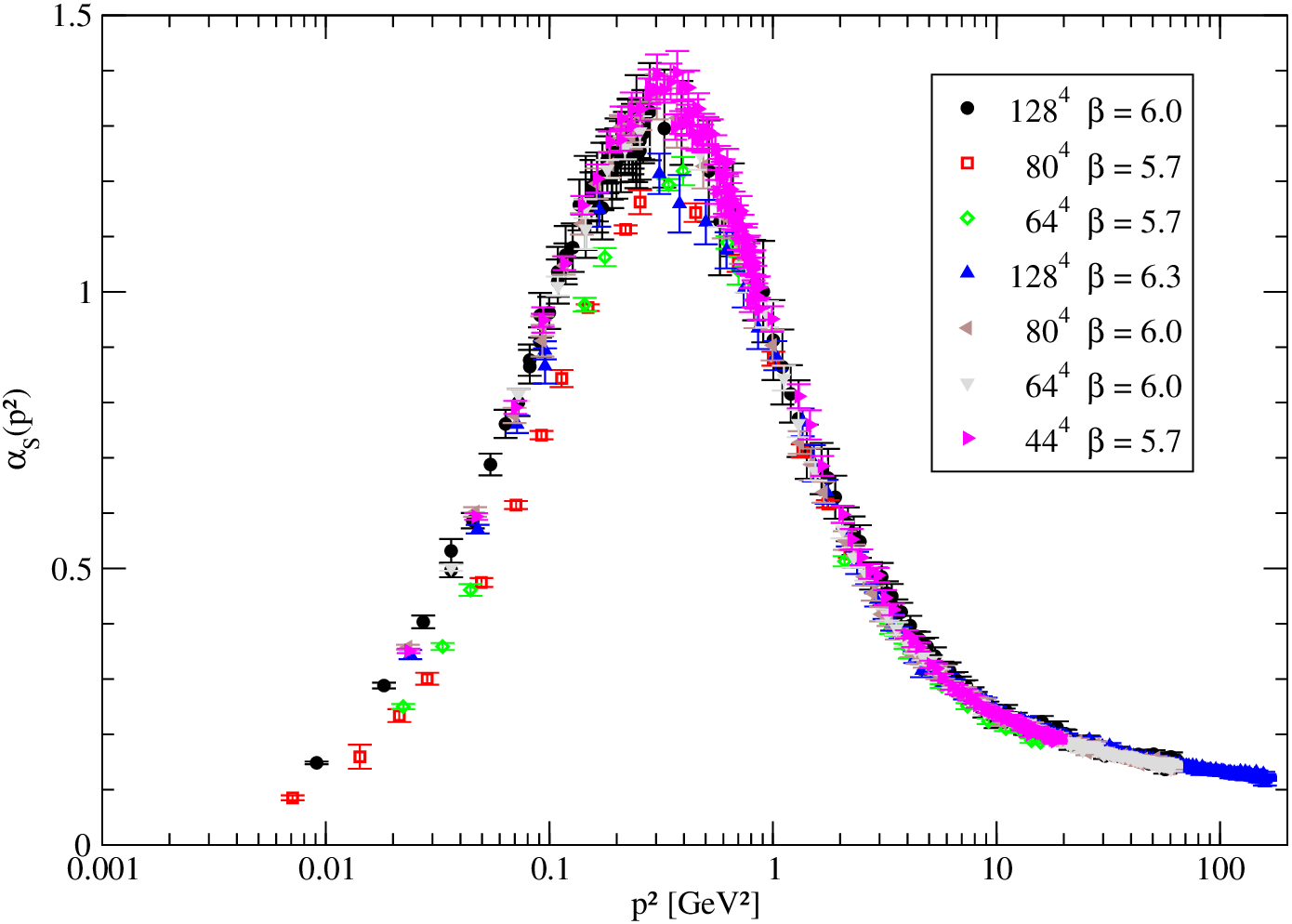} 
   \caption{Comparison of the results for the strong coupling computed from the simulations reported in Tab.~\ref{tab:setup} and Tab.~\ref{tab:berlinsetup}. }
   \label{fig:alpha_all}
\end{figure}

The comparison of the results for the strong coupling with those obtained by the 
Berlin-Moscow-Adelaide group can be seen in Fig.~\ref{fig:alpha_all}.
The differences between the two sets of simulations are
clearly seen for $p \lesssim 1$ GeV, with the estimations of~\cite{Bogolubsky:2009dc}
being smaller than those obtained in our simulations.
In fact, some dependence on the lattice volume can be seen by comparing the different $\beta=5.7$ data at low momenta.
The results of all the simulations become compatible for momenta above $\sim 1$ GeV, as already described in Sec. \ref{Sec:coupling}.

\section{Summary and Conclusions \label{Sec:fim}}

In this work we report the dependence of the lattice results for the gluon propagator, the ghost propagator and 
the strong coupling constant on the physical volume and lattice spacing used to simulate QCD. 
Our goal is to understand how precise one can compute these functions using lattice QCD simulations, modulo possible effects
associated with the presence of Gribov copies. 
In fact, the issue of the Gribov copies can be 
important to the calculation of the propagators and, possibly, 
the strong coupling constant \cite{Silva2004, Silva2007, Sternbeck2013,Sternbeck:2005tk}. 
However, due to the huge amount of computer time needed to study Gribov
copies effects in these large lattices, we are unable to disentangle 
such influence in our lattice results. Nevertheless, the observed dependence 
of the above mentioned functions on the lattice spacing and
physical volume is, as demonstrated by the results discussed here 
and~\cite{Oliveira:2012eh}, far from trivial and impact mainly in the low
momentum region.

In what concerns the gluon propagator the new simulations reproduce the behaviour already observed in~\cite{Oliveira:2012eh}.
The lattice data show essentially no dependence on the lattice physical volume for volumes above (6.5 fm)$^4$ for the full range of momenta
accessed. On the other hand, the infrared propagator reveals a non trivial dependence on the lattice spacing, with smaller lattice 
spacings favouring larger infrared propagators for momenta smaller than $\sim$1 GeV.

On the other hand,
the ghost propagator has a mild dependence on the lattice volume but, contrary to the gluon propagator, the simulations show that
this two point correlation function is suppressed at low momenta when the lattice spacing is decreased. We would like to point out
that, for the ghost propagator, the functional form (\ref{Eq:FitRenGhost}) which reproduces the perturbative one-loop result at high momentum
is able to describe the lattice data over a surprisingly wide range of momenta. Indeed, 
if one takes $\Lambda$ as fitting parameter, (\ref{Eq:FitRenGhost}) 
is able to fit the lattice data from momenta $\sim 1$ GeV up to the largest momenta simulated. If one sets
$\Lambda \sim \Lambda_{QCD} \sim 200$ MeV, the range of momenta described by (\ref{Eq:FitRenGhost}) starts from about
$\sim 2$ GeV and goes, again, up to the largest momentum available. We take this result as an indication that the ghost propagator follows
closely the perturbative propagator for momenta as small as $\sim 1$ GeV.

From Figs.~\ref{fig:gluon_prop_4GeV}
and~\ref{fig:ghost_dressing_4GeV} one can quantify how the propagators are modified by changing the lattice spacing.
For the gluon propagator one finds, for zero momentum,
a $\sim 10 \%$ order of magnitude effect by decreasing the lattice spacing from 0.18 fm down to
0.06 fm. In what concerns the ghost propagator, the change of the lattice spacing changes the propagator by $\sim 7 \%$ 
at the lower momenta available in our simulations.

The dependence of the strong coupling constant on the lattice spacing and physical volume is milder than for the propagators. Although the 
position of the maximum of $\alpha_s (p^2)$ seems to be independent of the both the lattice spacing and volume, the value of the strong
coupling constant seems to be suppressed as one approaches the continuum limit. As discussed in Sec.~\ref{Sec:coupling}, the value of
$\alpha_s (p^2)$ at the maximum is reduced by $\sim 15 \%$ for our largest value, when compared to the other calculations.

In Sec. \ref{Sec:compbma} our results are compared with those obtained using the largest physical volumes to simulate pure Yang-Mills
SU(3) theory~\cite{Bogolubsky:2009dc}. Such large volumes were achieved by relying on a large lattice spacing $a = 0.18$ fm. 
Our simulations and those performed by the Berlin-Moscow-Adelaide group give different answers for all the quantities considered here
at low momenta. Note, however, that at the qualitatively level the propagators and the strong coupling constant are similar.
Furthermore, looking at the renormalized data, see Figs.~\ref{fig:gluon_prop_4GeV} for the gluon and Fig.~\ref{fig:glue_prop_berlin_renormalised}
for the ghost data, both sets of propagators show no dependence or a very mild one on the physical volume.
The differences that are seen in the infrared for the two sets of simulations may be explained by different choices of the gauge fixing algorithm, i.e.
in principle it can be attributed to Gribov noise.
Indeed, it is well known that the choice of the maxima of $F_U[g]$ can change the propagators in the low momenta region.
A direct comparison of the Berlin-Moscow-Adelaide simulations and ours for $a = 0.18$ fm, suggests that the continuum gluon (ghost) propagator 
should be suppressed (enhanced) at low momenta.

In summary, our results show that the computation of the two point correlation 
functions on the lattice has a non-trivial dependence on the lattice spacing
and a mild dependence on the lattice volume for volumes above (6.5 fm)$^4$. 
Simulations performed with large lattice spacings, i.e. $a \gtrsim 0.18$ fm 
for pure Yang-Mills theory, are able to get the qualitative 
features of the propagators but introduce a measurable bias on the results 
at low momenta. The use of such large lattice spacings introduce also
strong lattice spacing effects for all momenta range, not show here, which 
are removed for momenta above $\sim 1$ GeV by performing
cuts on the momenta~\cite{Leinweber1998}. 
All the simulations discussed 
here use the Wilson action; certainly, improving the action may ameliorate
 the results in what concerns the dependence on the lattice spacing. However,
relying on improved actions requires revising all the calculation procedure.

\begin{acknowledgments}
The authors acknowledge the Laboratory for Advanced Computing at University of Coimbra for providing HPC computing resources 
(Milipeia, Centaurus and Navigator) that have contributed to the research results reported within this paper (URL http://www.lca.uc.pt). 
The authors also acknowledge computing resources provided by the Partnership for Advanced Computing in Europe (PRACE) initiative
under DECI-9 project COIMBRALATT and DECI-12 project COIMBRALATT2. 
P. J. Silva acknowledges support by F.C.T. under contract SFRH/BPD/40998/2007. 
The authors thank Balint Joo and Robert Edwards for helpful 
discussions concerning the use of Chroma, and Michael Pippig for the use of PFFT.
This work was partially supported by projects CERN/FP/123612/2011, CERN/FP/123620/2011 and PTDC/FIS/100968/2008, projects developed under initiative QREN financed by UE/FEDER through Programme COMPETE.
\end{acknowledgments}



\end{document}